# Clear evidence of charge conjugation and parity violation in K atoms from atomic permanent electric dipole moment experiments

Pei-Lin You[1]    Xiang-You Huang[2]
1. Institute of Quantum Electronics, Guangdong Ocean University, Zhanjiang Guangdong 524025, China.
2. Department of Physics, Peking University, Beijing 100871, China.

Quantum mechanics thinks that atoms do not have permanent electric dipole moment (EDM) because of their spherical symmetry. Therefore, there is no polar atom in nature except for polar molecules. The electric susceptibility $x_e$ caused by the orientation of polar substances is inversely proportional to the absolute temperature T while the induced susceptibility of atoms is temperature independent. This difference in temperature dependence offers a means of separating the polar and non-polar substances experimentally. Using special capacitor our experiments discovered that the relationship between $x_e$ of Potassium atom vapor and T is just $x_e=B/T$, where the slope B≈283(k) as polar molecules, but appears to be disordered using the traditional capacitor. Its capacitance C at different voltage V was measured. The C−V curve shows that the saturation polarization of K vapor has be observed when E≥$10^5$V/m and nearly all K atoms (more than 98.9％) are lined up with the field! The ground state neutral K atom is polar atom with a large EDM : $d_K$ >2.3×$10^{-9}$e.cm. The results gave clear evidence for CP (charge conjugation and parity) violation in K atoms. If K atom has a large EDM, why the linear Stark effect has not been observed? The article discussed the question thoroughly. Our results are easy to be repeated because the details of the experiment are described in the article.

**1.Introduction**   In order for an atom or elementary particle to possess a permanent electric dipole moment (EDM), time reversal (T) symmetry must be violated, and through the CPT theorem CP(charge conjugation and parity) must be violated as well[1]. The currently accepted Standard Model of Particle Physics predicts unobservable the dipole moments of an atom, therefore, EDM experiments are an ideal probe for new physics beyond the Standard Model. Experimental searches for EDMs can be divided into three categories: search for the neutron EDM[2], search for the electron EDM utilizing paramagnetic atoms, the most sensitive of which is done with Tl atoms(the result is $d_e$=[1.8±1.2 (statistical)±1.0 (systematic)]×$10^{-27}$ e.cm)[3], and search for an EDM of diamagnetic atoms, the most sensitive of which is done with $^{199}$Hg(the result is d(Hg)=-[1.06±0.49 (stat)±0.40 (syst)]×$10^{-28}$e.cm )[1,4]. Experiments to search for an EDM of atom began many decades ago, no large EDM has yet been found [1-5]. In all experiments, they measured microcosmic Larmor precession frequency of individual particle based on nuclear spin or electron spin. The search for an EDM consists of measuring the precession frequency of the particle in parallel electric and magnetic fields and looking for a change of this frequency when the direction of **E** is reversed relative to **B.** We now submit the article on the similar topic, however, with measuring macroscopic electric susceptibility($x_e$) of K vapor containing a large number of K atoms (the density N ≈$10^{20}$$m^{-3}$). Our experiments showed that ground state neutral K atom is polar atom with a large EDM. On the other hand, some evidence for CP violation beyond the Standard Model comes from Cosmology. Astronomical observations indicate that our Universe is mostly made of matter and contains almost no anti-matter. The first example of CP violation was discovered in 1964, but it has been observed only in the decays of the $K_o$ mesons. After 38 years, the BaBar experiment at Stanford Linear Accelerator Center (SLAC) and the Belle collaboration at the KEK laboratory in Japan announced the second example of CP violation in B mesons. "The results gave clear evidence for CP violation in B mesons. However, the degree of CP violation now confirmed is not enough on its own to account for the matter-antimatter imbalance in the Universe." "In the future, the BaBar experiment allows us to examine rarer processes and more subtle effects that will give us an even clearer understanding and may point us towards the processes which caused our universe to evolve into its current state." (SLAC Press Release July 23, 2002). So EDM experiments are now considered an ideal probe for evidence of new sources of CP violation. If an EDM is found, it will be compelling evidence for the existence of new sources of CP violation. **Our experimental results gave clear evidence for CP violation in K atoms. Few experiments in atomic physics have produced a result as surprising as this one.** This finding is a vital clue that an unknown factor is very likely at play in K atoms, such as Rb and Cs atoms[6,22]. This result is the product of eight years of intense



research and is a classic example of how understanding of our universe advances through atomic physics research [14]. The correctness of our theoretical motivation is supported by such five facts as the following.

①The shift in the energy levels of an atom in an electric field is known as the Stark effect. Normally the effect is quadratic in the field strength, but first excited state of the hydrogen atom exhibits an effect that is linear in the strength. This is due to the degeneracy of the excited state. This result shows that the hydrogen atom (the quantum number n=2 ) has very large EDM, $d_H=3ea_o=1.59\times 10^{-8}$e.cm ($a_o$ is Bohr radius)[7,8]. On the one hand, this EDM does not depend on the field intensity, hence it is not induce by the external field but is inherent behavior of the atom[7,8]. L.I. Schiff once stated that "Unperturbed degenerate states of opposite parities, as in the case of the hydrogen atom, can give rise to a permanent electric dipole moment"[7]. L.D. Landay also once stated that "The presence of the linear effect means that, in the unperturbed state, the hydrogen atom has a dipole moment"[8]. But on the other hand, the calculation of quantum mechanics tells us that unperturbed degenerate states of hydrogen atom(n=2) with zero EDM and has result $\langle \psi_{2lm} | er | \psi_{2lm} \rangle=0$, where $\psi_{2lm}$ are four wave functions of unperturbed degenerate states [8]. Due to the EDM of the hydrogen atom is responsible for the presence of linear Stark effect. A hydrogen atom(n=2) with zero EDM how responds to the external field and results in the linear Stark effect ? Quantum mechanics can not answer the problem [9-11]!

②In addition, the radius of the hydrogen atom of the first excited state is $r_H = 4a_o =2.12\times 10^{-8}$cm, it is almost the same as the radius of $^{199}$Hg ($r_{Hg}=1.51\times 10^{-8}$ cm) [12], but the discrepancy between their EDM is by some twenty orders of magnitude! How do explain this inconceivable discrepancy? The current theory can not answer the problem! The current theory thinks that in quantum mechanics there is no such concept as the path of an electron [8]. No one will give you any deeper explain of the inconceivable discrepancy. No one has found any more basic mechanism from which these results can be deduced! However, a hydrogen atom (n=2) has a nonzero EDM in the semi-classical theory of atom. The electron in a hydrogen atom (n=2) moves along a quantization elliptic orbit. We can draw a straight line perpendicular to the major axis of the elliptic orbit through the nucleus in the orbital plane. The straight line divides the elliptic orbit into two parts that are different in size. According the law of conservation of angular momentum, the average distance between the moving electron and the static nucleus is larger and the electron remains in the large part longer than in the small part. As a result, the time-averaged value of the electric dipole moment over a period is nonzero for the atom.

③The alkali atoms having only one valence electron in the outermost shell can be described as hydrogen-like atoms[13]. Since the quantum number of the ground state alkali atoms are n≥2 rather than n=1, in the Sommerfeld picture the valence electron moves along a highly elliptical orbit, the so-called diving orbits, approach the nucleus, as the excited state of the hydrogen atom. So we conjecture that the ground state neutral alkali atoms may have large EDM of the order of $ea_o$, but the actual result of the conjecture still needs to be tested by experiments [14].

④Quantum mechanics thinks that atoms do not have EDM because of their spherical symmetry. Therefore, there is no polar atom in nature. When atoms are placed in an electric field, they become polarized, acquiring *induced* electric dipole moments in the direction of the field. On the other hand, many molecules do have EDM. This molecule is called a polar molecule, such as $H_2O$, HCl, etc. Note that the susceptibility($x_e$) caused by the orientation of polar molecules is inversely proportional to the absolute temperature(T): $x_e = B/T$ while the induced susceptibility due to the distortion of electronic motion in atoms is temperature independent: $x_e=A$, where A and the slope B is constant, $x_e =C/C_o - 1$, $C_o$ is the vacuum capacitance and C is the capacitance of the capacitor filled with the material[15]. J.D. Jackson once stated that this difference in temperature dependence offers a means of separating the polar and non-polar substances experimentally [15].

⑤R.P. Feynman considered the orientation polarization of water vapor molecules [16]. He plotted the straight line from four experimental points: $x_e =A +B/T \approx B/T$, where A≈0 . Table 1 gives the four experimental data [17].

**Table 1 Experimental measurements of the susceptibility of water vapor at various temperatures**

| T(K) | Pressure(cm Hg) | $x_e$ |
|---|---|---|
| 393 | 56.49 | 0.004002 |
| 423 | 60.93 | 0.003717 |
| 453 | 65.34 | 0.003488 |
| 483 | 69.75 | 0.003287 |

From the ideal gas law, the average density of water vapor is $N = P/kT = (1.392\pm 0.002)\times 10^{25}$ m$^{-3}$. The electric susceptibility caused by the orientation of gaseous polar molecules is given by the expression

$$x_e = N d_o^2 /3kT \varepsilon_o \qquad (1)$$



where k is Boltzmann constant, $\varepsilon_o$ is the permittivity of free space, $d_o$ is the EDM of a molecule and the slope $B = N d_o^2 / 3k\varepsilon_o$. We work out the slope of the line $B = (1.50 \pm 0.04)$K and the EDM of a water molecule $d_{H2O} = (3k\varepsilon_o B / N)^{1/2} = (6.28 \pm 0.18) \times 10^{-30}$ C. m. The result in agreement with the observed value $d_{H2O} = 6.20 \times 10^{-30}$ C. m[18]. In addition, from $B \approx 0.94$ K the EDM of a HCl molecule can be deduced to be $d_{HCl} = 3.60 \times 10^{-30}$ C. m $\approx 0.43 ea_o$ [18]. The molecular electric dipole moments are of the order of magnitude of the electronic charge ($1.6 \times 10^{-19}$ C) multiplied by the molecular dimensions ($10^{-10}$ m), or about $10^{-30}$ coulomb-meters. **If K atom is the polar atom, a temperature dependence of the form $x_e = B/T$ should be expected in measuring the susceptibility $x_e$.**

**2. Experimental method and result** The first experiment: investigation of the relationship between the electric susceptibility $x_e$ of K vapor and absolute temperatures T using the traditional capacitor. The experimental apparatus was a closed glass container with two concentric metal tubes, respectively showed by $A_1B_1$ and $A_2B_2$ in Fig.1. The tube $A_1B_1$ and $A_2B_2$ build up a cylindrical capacitor, where H=R-r<<r. The capacitance was measured by a digital meter with a resolution of 0.1pF, an accuracy of 0.2% and the surveying voltage V=1.0 volt. When the container is empty, it is pumped to vacuum pressure P $\leq 10^{-6}$ Pa for 20 hours. The measured capacitances is $C_0 = (54.2 \pm 0.1)$ pF. Next, a small amount of K material with high purity was put in the container. The container was again pumped to P $\leq 10^{-6}$ Pa, then it was sealed. Thus we obtain a cylindrical capacitor filled with K vapor at a fixed density (in more detail see arXiv. 0809.4767). Now, its capacitances was $C_t = (78.6 \pm 0.1)$pF ($x_e > 0.4$). We put the capacitor into a temperature-control stove, raise the temperature of the stove very slow in the temperatures 30℃—210℃. By measuring the electric susceptibility $x_e$ of K vapor at different temperature T, the experimental results are shown in Fig.2. The experimental curve shows that the relationship between $x_e$ of K vapor and T appears to be disordered using the traditional capacitor.

The second experiment: investigation of the relationship between the electric susceptibility $x_e$ of K vapor and absolute temperatures T using the repackage capacitor. The apparatus of the second experiment was the same as the preceding one but the capacitor was repackaged. The longitudinal section of the repackage capacitor is shown in Fig 3. The external surface of the container was plated with silver which showed by AB. The actual cylindrical capacitor is consists of the metal tube $A_1B_1$ and the silver layer AB rather than the tube $A_2B_2$. This cylindrical capacitor is connected in series by two capacitors. One is called C' and contains K vapor with thickness H. The other is called C" and contains the glass medium with thickness △=1.2mm. The capacitance of the capacitor can be written as C=C'C"/(C'+C") or C'=CC"/(C"–C), where C and C" can be directly measured in the experiment. The radiuses of AB and $A_1B_1$ are shown respectively by **R** and **r**. Since R-r=H<<r, the cylindrical capacitor could be approximately regarded as a parallel-plane capacitor. The capacitance C was still measured by the digital meter and $C_0 = (74.0 \pm 0.2)$pF (the plate area S=$(4.8 \pm 0.2) \times 10^{-2}$ m$^2$, the plate separation H=$(5.8 \pm 0.2)$mm). The capacitor filled with K vapor and the density N of K vapor remained fixed. We put the repackage capacitor into the stove and raise the temperature of the stove slow. The capacitances have been measured at different temperatures T, the experimental results are shown in Fig.4. Table 2 gives the experimental data. The experiment to measure C" is very easy. We can make a cylindrical capacitor of glass medium with thickness △ and put it in the stove. Notice that the dielectric constant of the glass medium vary from 6 to over 12 in the temperatures range 363K—463K.

**Table 2 Experimental measurements of the electric susceptibility of K vapor at various temperatures**

|   | T(K) | 1/T | C(pF) | C"(pF) | C'(pF) | $x_e$ |
|---|------|-----|-------|--------|--------|-------|
| 1 | 363 | 0.002755 | 111.5 | 1668 | 119.5 | 0.6149 |
| 2 | 385 | 0.002597 | 109.4 | 1946 | 115.9 | 0.5664 |
| 3 | 403 | 0.002481 | 108.2 | 2228 | 113.7 | 0.5367 |
| 4 | 425 | 0.002353 | 107.0 | 2785 | 111.3 | 0.5037 |
| 5 | 443 | 0.002257 | 105.3 | 3064 | 109.0 | 0.4736 |
| 6 | 463 | 0.002160 | 103.6 | 3340 | 106.9 | 0.4448 |

Where **C** ---the total capacitance, **C"**----the capacitance of the glass medium, **C'**----the capacitance of the K vapor, the electric susceptibility of K vapor $x_e = (C' - C_0)/C_0$ and the slope B= $\triangle x_e / \triangle (1/T)$.

We obtain the electric susceptibility of K vapor $x_e = A + B/T \approx B/T$, where the intercept A at 1/T= 0 represents the susceptibility due to induced polarization at this density and $A \approx 0$, while the average slope $B \approx 283$(K), where: $B_{12}$=307(K), $B_{13}$=285(K), $B_{14}$=277(K), $B_{15}$=284(K), $B_{16}$=286(K), $B_{23}$=256(K), $B_{24}$=257(K), $B_{25}$=273(K), $B_{26}$=278(K), $B_{34}$=258(K), $B_{35}$=282(K), $B_{36}$=286(K), $B_{45}$=314(K), $B_{46}$=305(K), $B_{56}$=297(K) and $B_{ij} = [\triangle x_e / \triangle (1/T)]_{ij}$.

The orientation polarization of K atoms in a weak field has been observed, as polar molecules. This experimental



result exceeded all expert's expectation and the most reasonable explanation is that K atom is polar atom.

The third experiment: measuring the capacitance of K vapor at various voltages (V) under a fixed density N and the temperature $T_1$. The apparatus was the same as the second experiment($C_0$=(74.0±0.2)pF) and the method is shown in Fig. 5. C was the capacitor filled with K vapor to be measured and kept at $T_1$ =(328±0.5)K. $C_d$ was used as a standard capacitor. Two signals $V_c(t)=V_{co} \cos \omega t$ and $V_s(t)=V_{so} \cos \omega t$ were measured by a digital oscilloscope having two lines. The two signals had the same frequency and always the same phase at different voltages when the frequency is greater than a certain value. It indicates that capacitor C filled with K vapor was the same as $C_d$, a pure capacitor without loss. From Fig.5, we have $(V_s-V_c)/V_c=C/C_d$ and $C=(V_{so}/V_{co}-1)C_d$. In the experiment $V_{so}$ can be adjusted from zero to 800V. The measured result of the capacitance C at various voltages was shown in Fig.6. When $V_{co}=V_1=$ 0.5v, $C_1$=(138.0±0.2)pF is approximately constant. With the increase of voltage, the capacitance decreases gradually. When $V_{co}=V_2 \geq (600\pm5)$v, $C_2 \leq (76.0\pm0.2)$pF, it approaches saturation. If all the dipoles in a gas turn toward the direction of the field, this effect is called the saturation polarization. The C—V curve shows that the saturation polarization of the K vapor is obvious when $E=V_2/H \geq 10^5$v/m.

## 3. Discussion

①Our experimental apparatus is a closed glass container. In all experiments, when the capacitor is empty, it is pumped to vacuum pressure P $\leq 10^{-6}$ Pa for 20 hours. The aim of the operation is to remove carefully impurities, such as oxygen, absorbed on the inner walls of the container. In addition, the purity of K material exceeded 99.95 %. **Actual result showed that the K vapor filled the capacitors is present in atomic form, not the dimer.**

K vapor will cause the glass surface which contacts with the vapor to be weak conductivity. What influence on the experimental result will be produced by the conductivity? Because the conductive film is a surface of equal potentials for the cylindrical capacitor, it has no effect to the capacitance of the capacitor.

②The electric susceptibility of polar molecules is of the order of $10^{-3}$ for gases at NTP and $10^0$ (or one) for solids or liquid at room temperature [15]. Experimentally, typical values of the susceptibility are 0.0046 for HCl gas, 0.0126 for water vapor, 5-10 for glass, 3-6 for mica[18]. When the surveying voltage V=1.0 volt using the digital meter, our experiments showed that the susceptibility of K vapor $x_e=C/C_0-1 > 0.40$ ( where the density N≈ $10^{20}$m$^{-3}$)! **We have seen new phenomenon not explained by current theories.**

③The electric susceptibility caused by the orientation of gaseous polar molecules is[19]

$$x_e = N d_o L(a)/ \varepsilon_o E \qquad (2)$$

where $\varepsilon_o$ is the permittivity of free space, N is the number density of molecules, $d_o$ is the EDM of a molecule. The mean value of $\cos \theta$ : $<\cos \theta> = \mu \int_o^\pi \cos \theta \exp(d_o E \cos \theta /kT) \sin \theta d\theta = L(a)$, where $a = d_o E /kT$, $\mu$ is a normalized constant, $\theta$ is the angle between $\mathbf{d_o}$ and $\mathbf{E}$. $L(a) = [(e^a + e^{-a})/(e^a - e^{-a})] - 1/a$ is called the Langevin function, where a<<1, $L(a) \approx a/3$ and a>>1, $L(a) \approx 1$[19]. When the saturation polarization appeared, $L(a) \approx 1$ and this will happen only if a>>1[19]. R.P. Feynman once stated that *"when a filed is applied, if all the dipoles in a gas were to line up, there would be a very large polarization, but that does not happen"* [16]. So, no scientist has observed the saturation polarization effect of any gaseous dielectric till now! **The saturation polarization of K vapor in ordinary temperatures is an entirely unexpected discovery.**

④Due to the induced dipole moment of K atom is $d_{int}=G\varepsilon_o E$, where G=43.4×$10^{-30}$m$^3$ [20], the most field strength is $E_{max} \leq 10^5$v/m in the experiment, then $d_{int} \leq 3.9 \times 10^{-35}$ C.m can be neglected. We obtain

$$x_e = N d L(a)/ \varepsilon_o E \qquad (3)$$

where d is the EDM of an K atom and N is the number density of K atoms. **L(a)= $<\cos \theta>$ is the ratio of K atoms which lined up with the field in the total atoms.** Note that $x_e = \varepsilon_r - 1 = C/C_o - 1$, where $\varepsilon_r$ is the dielectric constant, $E=V/H$ and $\varepsilon_o = C_o H/S$, leading to

$$C - C_o = \beta L(a)/a, \qquad (4)$$

**This is the polarization equation of K atoms**, where $\beta = S N d^2/kTH$ is a constant. Due to a=dE/kT= dV/kTH, **we obtain the formula of atomic EDM**

$$d_{atom} =(C - C_o)V / L(a)SN \qquad (5)$$

In order to work out **L(a)** and **a** of the third experiment, note that when $V_1$=0.5 volt, a<<1 and $L(a) \approx a/3$, $C_1 - C_0 = \beta/3$ and $\beta$=192pF. When $V_2$= 600 volt, a>>1 and $L(a) \approx 1$, $C_2 - C_0 = L(a_2) \beta /a_2 \approx \beta /a_2$. When a → ∞, $x_e = 0$ and the dielectric constant of the K vapor is the same as vacuum! We work out $a_2 \approx \beta/(C_2 - C_0)$=96, $L(a_2) \approx$ L(96)=0.9896, $a_2 = \beta L(a_2)/(C_2 - C_0)$=95 and $L(a_2)$=L(95)=0.9895. Then $a_1 = a_2 V_1/V_2$=0.07917<<1 and $L(a_1) \approx a_1/3$=0.02639. $L(a_2)$= 0.9895 means that more than 98.9% of K atoms have be oriented in the direction of the field



when $V_2$ =600volt or E=$V_2$/H≈$10^5$V/m. L($a_1$) =0.02639 means that only 2.64％ of K atoms have be oriented in the direction of the field when $V_1$=0.5volt or E≈86V/m. The formula of saturated pressure of K vapor is P=$10^{7.183-4434.3/T}$, the effective range of the formula is 533K ≤T ≤1033K [12]. The saturated pressure of K vapor at 533K is P=0.07303 psi＝503.5 Pa. From the ideal gas law N= P/kT, the density of K vapor at 533K under the saturated pressure is

$$N_1 =503.5 /1.38 \times 10^{-23} \times 533 = 6.84 \times 10^{22} \text{ m}^{-3} \qquad (6)$$

Because the density N of K vapor is unknown in the experiments, we can estimate with certainly that $10^{20}$ m$^{-3}$< N <$10^{21}$ m$^{-3}$. Note that N < $N_{max}$= $N_1$ /10, i.e. N <6.84×$10^{21}$ m$^{-3}$, we obtain the smallest EDM limit of an K atom

$$d_K >(C_1 － C_0 ) V_1 / L(a_1)S\, N_{max} = (C_2 － C_0 ) V_2 / L(a_2)S\, N_{max} = 2.3 \times 10^{-9} \text{e.cm} \qquad (7)$$

**Although above calculation is simple, but no physicist completed the calculation up to now!**

⑤The formula $d_{atom}$ =(C－$C_o$)V/L(a)SN can be justified easily. The magnitude of the dipole moment of an K atom is d = e r. N is the number of K atoms per unit volume. L(a) is the ratio of K atoms lined up with the field in the total number. Suppose that the plates of the capacitor have an area S and separated by a distance H, the volume of the capacitor is SH. When an electric field is applied, the K atoms tend to orient in the direction of the field as dipoles. On the one hand, the change of the charge of the capacitor is △Q=(C－$C_0$ )V. On the other hand, when the K atoms are polarized by the orientation, the total number of K atoms lined up with the field is SHNL(a).The number of layers of K atoms which lined up with the filed is H/r. Because inside the K vapor the positive and negative charges cancel out each other, the polarization only gives rise to a net positive charge on one side of the capacitor and a net negative charge on the opposite side. Then the change △Q of the charge of the capacitor is △Q=SHN L(a)e / (H/r)= SN L(a)d = (C－$C_0$ )V, so the EDM of an K atom is **d = (C－$C_0$ )V/ SN L(a).**

⑥If K atom has a large EDM, why has not been observed in other experiments? This is an interesting question. In Eq.(4) let the function f (a)= L(a)/a and from f ″ (a)=0, we work out the knee of the function L(a)/a at $a_k$=1.9296813≈1.93. Corresponding knee voltage $V_k$=$V_2 a_k$/$a_2$ ≈12.2volt, knee field $E_k$=$V_k$/H≈2.1×$10^3$v/m. By contrast with the curve in Fig.6, it is clear that our polarization equation is valid. The third experiment showed that the saturation polarization of K vapor is easily observed. When the saturation polarization of K vapor occurred (V ≥600 volt), nearly all K atoms (more than 98％) are lined up with the field, and C≈$C_o$ ($C_o$ is the vacuum capacitance)! So only under the very weak field (E<$E_k$ i.e. E<2.1×$10^3$V/m), above phenomenon of K atom can be observed. **Regrettably, nearly all scientists in this field disregard the very important problem.**

⑦As a concrete example, let us treat the linear Stark shifts of the hydrogen(n=2) and K atom. Notice that the fine structure of the hydrogen (n=2) is only 0.33 cm$^{-1}$, and therefore the fine structure is difficult to observe [21]. The linear Stark shifts of the energy levels is proportional to the field strength: △W= d.E=3e$a_o$E=1.59×$10^{-8}$ E e.cm. When E=$10^5$V/cm, △W=1.59×$10^{-3}$ eV, this corresponds to a wavenumber of 12.8 cm$^{-1}$. For the Hα lines of the Balmer series( $\lambda$ = 656.3 nm, and the splitting is only 0.014 nm can be neglected)[21], the linear Stark shifts is △$\lambda$ = △W$\lambda^2$/hc =0.55 nm. It is so large, in fact, that the Stark shift of the lines is easily observed [21]. Because the most field strength for K atoms is $E_{max}$=$10^5$V/m=$10^3$V/cm, if K atom has five times as large EDM d=5×2.30 ×$10^{-9}$ e.cm=1.15×$10^{-8}$ e.cm, and the most splitting of the energy levels of K atoms △$W_{max}$= d.$E_{max}$=1.15×$10^{-5}$ eV. This corresponds to a wavenumber of 9.3×$10^{-2}$ cm$^{-1}$. On the other hand, the observed values for a line pair of the first primary series of K atom( Z=19, n=4) are $\lambda_1$=769.90 nm and $\lambda_2$=766.49 nm[12]. The magnitude of the linear Stark shift of K atoms is about △$\lambda$ = △W ($\lambda_1$ + $\lambda_2$)$^2$ / 4hc = 0.0055nm. **It is so small, in fact, that a direct observation of the linear Stark shifts of K atom is not possible.**

The striking features of our experiments are the data were reliable and the measuring process can be easily repeated in any university or laboratory because the details of the experiments are described in the paper. Our experimental apparatus are still kept, we welcome anyone who is interested in the experiments to visit and examine it. A detail explanation of how fill with K, Rb or Cs vapor at a fixed density in a vacuum environment as indicated in Ref. 22.

**References**

1. M. V. Romalis, W .C. Griffith, J.P. Jacobs and E N. Fortson, Phys. Rev. Lett. **86,** 2505 (2001)
2. P. G. Harris *et al*., Phys. Rev. Lett. **82,** 904 (1999)
3. E. D. Commins, S. B. Ross, D. DcMille, and B.C. Regan, Phys .Rev. A **50,** 2960 (1994)
4. J. P. Jacobs, W. M. Klipstein, S. K. Lamoreaux, B. R. Heckel, and E. N. Fortson, Phys. Rev. A **52,** 3521 (1995)
5. Cheng Chin, Veronique Leiber, Vladan Vuletic, Andrew J. Kerman, and Steven Chu , Phys. Rev. A **63**, 033401(2001)
6. H. R. Quinn, and M. S. Witherell, Sci. Am., Oct. 1998 PP76-81
7. L.I. Schiff , Quantum Mechanics (New York McGraw-Hill Book Company 1968)P488, PP252-255
8.L.D.Landay, and E.M.Lifshitz, Quantum Mechanics(Non-relativistic Theory) (Beijing World Publishing Corporation 1999)P290,P2
9. L. E. Ballentine, Quantum Mechanics a modern development (World Scientific Publishing Co, Pte Ltd 1998)P286, P373
10. L.E.Ballentine, Yumin Yang ,and J.P.Zibin, Phys. Rev. A. **50** 2854 (1994)
11. Xiang-You Huang, Phys. Lett. A **121** 54 (1987)
12. J.A. Dean, Lange's Handbook of Chemistry (New York: McGraw-Hill ,Inc 1998)Table 4.6,5.3,5.8 and 7.19

13. W. Greiner, Quantum Mechanics an introduction(Springer-verlag Berlin/Heidelberg 1994)PP254-258, P213.
14. Pei-Lin You, Study on occurrence of induced and permanent electric dipole moment of a single atom, J. Zhanjiang Ocean Univ. Vol.20 No.4, 60 (2000) (in Chinese)
15. J.D. Jackson, Classical Electrodynamics (John Wiley & Sons Inc. Third Edition1999), PP162-165, P173
16. R.P. Feynman, R.B. Leighton, and M. Sands, The Feynman lectures on physics Vol.2( Addison-Wesley Publishing Co. 1964), P11-5,P11-3.
17. Sänger, Steiger, and Gächter, Helvetica Physica Acta 5, 200(1932).
18. I.S. Grant, and W.R. Phillips, Electromagnetism. (John Wiley & Sons Ltd. 1975), PP57-65
19. C.J.F. Bottcher, Theory of Electric Polarization (Amsterdam: Elsevier 1973), P161
20. D.R. Lide, Handbook of Chemistry and Physics. (Boca Raton New York: CRC Press 1998), 10.201-10.202
21. H.Haken, and H.C.Wolf, The Physics of Atoms and Quanta. (Springer-Verlag Berlin Heidelberg 2000),P195,PP171-259
22. Pei-Lin You and Xiang-You Huang , arXiv.0809.4767 and 0810.0770.

**Acnowledgement**    The authors thank to our colleagues Zhao Tang , Rui-Hua Zhou, Zhen-Hua Guo, Ming- jun Zheng, Xue-ming Yi, , Xing Huang, and Engineer Jia You for their help in the work.

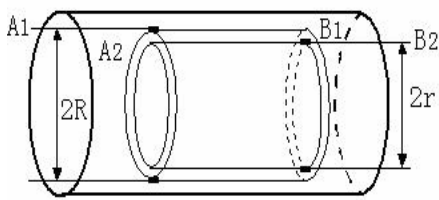
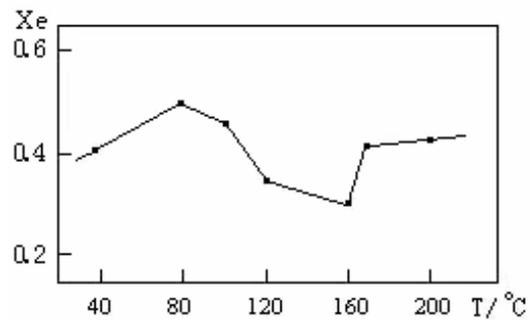

**Fig.1** The traditional capacitor was a closed glass container with two concentric metal tubes. The glass container filled with K vapor. The tube **A₁B₁** and **A₂B₂** build up a cylindrical capacitor, where H=R-r<<r.

**Fig.2** The experimental curve shows that the relationship between the electric susceptibility $x_e$ of K vapor and temperatures T appears to be disordered using the traditional capacitor.

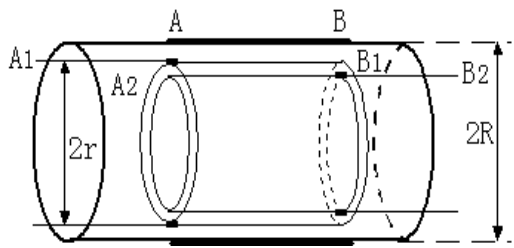

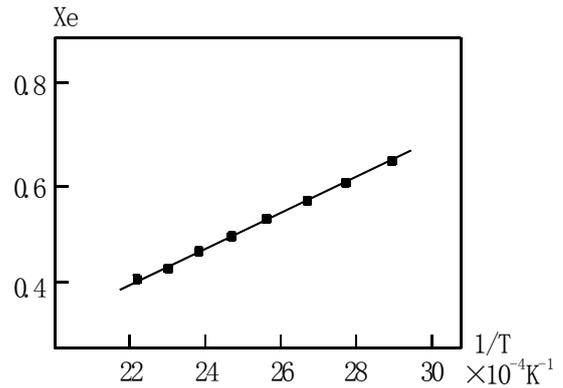

**Fig.3** The schematic diagram is a longitudinal section of the repackage capacitor filled with K vapor. The silver layer **AB** and the metal tube **$A_1B_1$** build up a cylindrical capacitor, where H=R-r<<r.

**Fig.4** The relationship between the electric susceptibility $x_e$ of K vapor and absolute temperatures T is $x_e$=B/T using the repackage capacitor, where the slope B≈283(K).

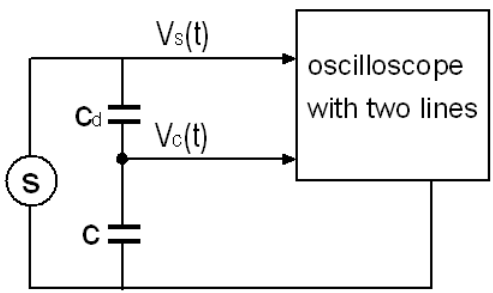

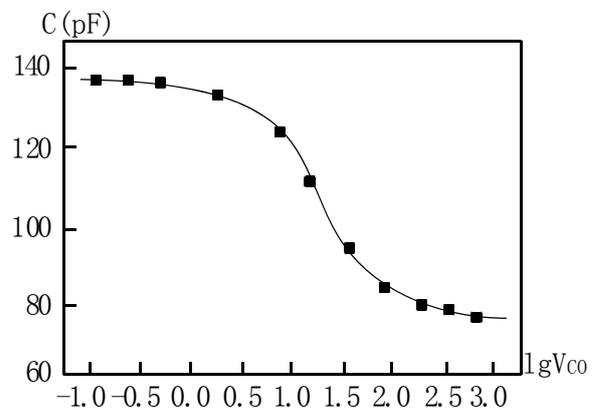

**Fig.5** The diagram shows the experimental method, in which C is capacitor filled with K vapor to be measured and $C_d$ is a standard one.
$V_s(t)=V_{so}\cos\omega t$    $V_c(t)=V_{co}\cos\omega t$

**Fig.6** The experimental curve shows that the saturation polarization effect of the K vapor is obvious when E≥$10^5$v/m.